# Is multiplexed off-axis holography for quantitative phase imaging more spatial bandwidth-efficient than on-axis holography?

GILI DARDIKMAN AND NATAN T. SHAKED*

*Department of Biomedical Engineering, Faculty of Engineering, Tel Aviv University, Tel Aviv 69978, Israel*
*Corresponding author: nshaked@tau.ac.il*

**Digital holographic microcopy is a thriving imaging modality that attracts considerable research interest due to its ability to not only create excellent label-free contrast, but also supply valuable physical information regarding the density and dimensions of the sample with nanometer-scale axial sensitivity. Three basic holographic recording geometries currently exist, including on-axis, off-axis and slightly off-axis holography, each of them enabling a variety of architectures in terms of bandwidth use and compression capacity. Specifically, off-axis holography and slightly off-axis holography allow spatial hologram multiplexing, enabling compressing more information into the same digital hologram. In this paper, we define an efficiency score used to analyze the various possible architectures, and compare the signal-to-noise ratio and mean squared error obtained using each of them, determining the optimal holographic method. © 2018 Optical Society of America**

http://dx.doi.org/10.1364/AO.99.099999

## 1. Introduction

Digital holographic microcopy yields excellent label-free contrast, even for samples that appear transparent in standard microscopy, while also supplying valuable physical information regarding the density and thickness of the sample with nanometer-scale axial sensitivity; this is possible due to an interferometric recording, able to capture the phase difference between two wavefronts. The phase difference is a valuable physical quantity [1-3], which quantifies the extent to which light was delayed when transpassing one path relative to another, and is proportional to the optical path delay (OPD) as follows:

$$\varphi(x,y) = \frac{2\pi}{\lambda} OPD(x,y), \qquad (1)$$

where λ is the illumination wavelength.

An image hologram captures the phase difference between a beam passing through a sample (sample beam) and a beam that does not (reference beam), by recording their interference pattern (which is the digital hologram, or interferogram) [3-6].

Three basic holographic recording geometries currently exist: on-axis, off-axis, and slightly off-axis holography, each of them enabling a variety of architectures in terms of bandwidth use and compression capacity. In Section 2 we review these methods. Then, in Section 3, we discuss the most efficient architectures for each of the recording geometries, and analyze them mathematically. In Section 4, we present a numerical simulation quantifying the quality of the reconstructed image in the presence of shot noise for the various holographic architectures discussed in Section 3. Finally, in Section 5, we conclude the discussion with practical thoughts regarding the optimal holographic method.

## 2. Principles of holographic phase imaging

### A. On-axis holography

In on-axis holography [3-6], the two interfering beams are projected onto the camera at the same angle, such that the digital hologram recorded by the camera is given by the following expression:

$$\begin{aligned}I_{\text{on-axis}}(x,y) &= |E_s(x,y) + E_r|^2 = \\ &|E_s(x,y)|^2 + |E_r|^2 + \\ &|E_s(x,y)||E_r|\exp[j \cdot \varphi_s(x,y)] + \\ &|E_s(x,y)||E_r|\exp[-j \cdot \varphi_s(x,y)],\end{aligned} \qquad (2)$$

where $j$ denotes the imaginary unit, $E_s(x,y)$ and $E_r$ are the sample and reference complex waves, respectively (the latter assumed to be of constant amplitude and phase), and $\varphi_s(x,y)$ is the phase difference between them.

The first two terms in Eq. (2) represent the sample and reference intensities, also referred to as zero orders, DC components, or auto-correlation terms. The last two terms in Eq. (2) are the complex conjugate cross-correlation (CC) terms, each containing the complex wavefront of the sample (amplitude and phase profiles).

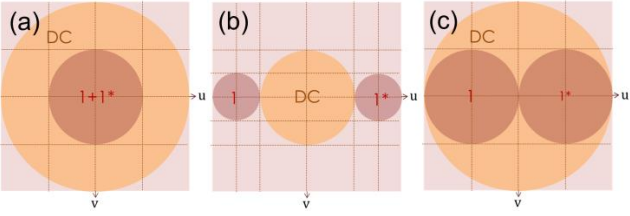

Fig. 1. Schematic illustrations of the SFD for the three typical holographic recording geometries. (a) On-axis holography. (b) Off-axis holography. (c) Slightly off-axis holography. DC denotes the auto-correlation terms, illustrated by an orange circle. The red circles illustrate the CC terms, where the coinciding complex conjugate CC terms are denoted by a number and an asterisk.

In order to isolate one of the CC terms, enabling the reconstruction of the phase difference $\varphi_s(x, y)$, at least three phase-shifted digital holograms are needed [6, 7], as can be achieved using wave plates or piezo mirrors in the path of either the sample or the reference beam. Once three such phase shifted holograms are acquired, each with 120° relative delay, we have three equations with three unknown variables – the DC terms, the positive CC term ($CC_1$) and the negative CC term ($CC_{-1}$) – allowing the isolation of each of the CC terms using the three measurements, as followed:

$$CC_1 = I_1 + I_2 \cdot \exp\left(\frac{2\pi j}{3}\right) + I_3 \cdot \exp\left(\frac{-2\pi j}{3}\right) \tag{3}$$

where $I_1$ is the hologram acquired with no phase shift, $I_2$ is the hologram acquired with a $2\pi/3$ phase shift, and $I_3$ is the hologram acquired with a $-2\pi/3$ phase shift [7].

A schematic illustration of the spatial-frequency domain (SFD) of an on-axis hologram is provided in Fig. 1(a).

**B. Off-axis holography**

In off-axis holography [6, 8-10], one of the interfering beams is tilted relative to the other beam at a small angle, creating a linear phase shift that allows separation of the field intensity from the two complex-conjugate wavefront terms in the SFD; this enables reconstruction of the complex wavefront from a single off-axis digital hologram. An off-axis hologram recorded by a camera can be mathematically expressed as follows:

$$\begin{aligned}I_{\text{off-axis}}(x, y) &= \left|E_s(x,y) + E_r \cdot \exp\left[j \cdot \varphi_o(x, y)\right]\right|^2 = \\ &\quad |E_s(x,y)|^2 + |E_r|^2 \\ &\quad + |E_s(x, y)||E_r|\exp\{j \cdot [\varphi_s(x, y) - \varphi_o(x, y)]\} \\ &\quad + |E_s(x, y)||E_r|\exp\{-j \cdot [\varphi_s(x, y) - \varphi_o(x, y)]\},\end{aligned} \tag{4}$$

where $\varphi_o(x, y)$ is the off-axis phase, induced by titling one of the beams (the reference beam here) in respect to the other. The off-axis phase is given by:

$$\varphi_o(x, y) = \frac{2\pi}{\lambda}[x \sin(\theta_x) + y \sin(\theta_y)], \tag{5}$$

where $\theta_x$ is the angle between the projection of the reference wave illumination direction on the x-z plain and the z axis, and $\theta_y$ is the angle between the projection of the reference wave illumination direction on the y-z plain and the z axis, assuming light propagates in the z direction. This linearly space-dependent phase is translated into a shifted Dirac delta function in the SFD, causing the CC terms multiplying it to shift from the center. This can be seen in Fig. 1(b) and in the following mathematical expression:

$$\begin{aligned}FT\{I_{\text{off-axis}}(x, y)\} &= \\ FT&\left\{|E_s(x,y)|^2\right\} + |E_r|^2 \cdot \delta(u,v) \\ &+ |E_r| \cdot FT\{|E_s(x, y)|\exp[j \cdot \varphi_s(x, y)]\} \\ &\quad *\delta\left[u + \frac{2\pi}{\lambda} \cdot \sin(\theta_x), v + \frac{2\pi}{\lambda} \cdot \sin(\theta_y)\right] \\ &+ |E_r| \cdot FT\{|E_s(x, y)|\exp[-j \cdot \varphi_s(x, y)]\} \\ &\quad *\delta\left[u - \frac{2\pi}{\lambda} \cdot \sin(\theta_x), v - \frac{2\pi}{\lambda} \cdot \sin(\theta_y)\right],\end{aligned} \tag{6}$$

where $(u, v)$ are the spatial-frequency coordinates and $\delta(u, v)$ is the Dirac delta function. As can be seen from Eq. (6), the DC terms remain centered in the SFD, while the CC terms are linearly shifted to opposite spatial frequencies according to their tilt angles, due to the convolution with shifted delta functions. The relation between the tilt angle and the desired shift can be formulated as follows:

$$\begin{aligned}u_0 &= \frac{2\pi}{\lambda} \cdot \sin(\theta_x) \Rightarrow \theta_x = \sin^{-1}\left(\frac{\lambda}{2\pi} \cdot u_0\right), \\ v_0 &= \frac{2\pi}{\lambda} \cdot \sin(\theta_y) \Rightarrow \theta_y = \sin^{-1}\left(\frac{\lambda}{2\pi} \cdot v_0\right),\end{aligned} \tag{7}$$

where $u_0$ and $v_0$ express the shift on the $u$ and $v$ axes in the SFD, respectively, expressed as fractions of the respective cutoff angular frequencies $\omega_{c,u}$ and $\omega_{c,v}$ [defined in Eq. (13)]; for example, in order to create the architecture presented in Fig. 1(b) we should choose $u_0 = 0.75\omega_{c,u}$, $v_0 = 0$.

For optically recorded holograms, the parameters of the optical setup dictate the SFD layout. A higher cutoff frequency can be obtained by using a smaller sampling unit. Equivalently, a lower SFD occupancy by the CC terms can be obtained by increasing the magnification of the optical system. However, the former is limited by the camera pixel size, and the latter comes at the cost of wasting camera pixels and decreasing the imaged field of view (FOV), or, alternatively, decreasing the framerate of the camera (since using a larger FOV reduces the maximal framerate). Thus, in general, we aim for the SFD occupancy to be as small as possible.

Nevertheless, if the spatial bandwidth is too narrow, the shifted CC terms may exceed the cutoff frequency and overlap with themselves in the SFD due to the cyclic property of the discrete Fourier transform (DFT), causing information loss. Thus, higher minimal sampling requirements must be used for off-axis holography relative to on-axis holography, potentially leading to wasteful bandwidth use.

This empty space in the SFD can be used for compressing additional information to a single hologram containing the same number of pixels as in each of the original holograms, by effectively placing additional non-overlapping CC terms in the SFD. This procedure is called hologram spatial-frequency multiplexing, or multiplexed holography [11,12]. The non-overlapping CC terms can contain additional information on the sample, such as different temporal events, different imaging areas, different color images, etc.

An ideal multiplexed hologram can be expressed by the simple summation of N off-axis holograms with different off-axis angles, given by:

$$I_{\text{multiplexed}}(x, y) = \sum_{k=1}^{N} \left| E_{s,k}(x,y) + E_{r,k} \cdot \exp\left[ j \cdot \varphi_{o,k}(x, y) \right] \right|^2. \quad (8)$$

The SFD of this ideal multiplexed hologram can be mathematically formulated as follows:

$$FT\{I_{\text{multiplexed}}(x, y)\} = \sum_{k=1}^{N} \left[ FT\left\{ |E_{s,k}(x,y)|^2 \right\} + |E_{r,k}|^2 \cdot \delta(u,v) \right]$$
$$+ \sum_{k=1}^{N} |E_{r,k}| \cdot FT\left\{ |E_{s,k}(x,y)| \exp\left[ j \cdot \varphi_{s,k}(x,y) \right] \right\}$$
$$* \delta\left[ u + \frac{2\pi}{\lambda} \cdot \sin(\theta_{x,k}), v + \frac{2\pi}{\lambda} \cdot \sin(\theta_{y,k}) \right] \quad (9)$$
$$+ \sum_{k=1}^{N} |E_{r,k}| \cdot FT\left\{ |E_{s,k}(x,y)| \exp\left[ -j \cdot \varphi_{s,k}(x,y) \right] \right\}$$
$$* \delta\left[ u - \frac{2\pi}{\lambda} \cdot \sin(\theta_{x,k}), v - \frac{2\pi}{\lambda} \cdot \sin(\theta_{y,k}) \right],$$

such that all DC terms remain centered in the frequency domain, while the multiple CC pairs are shifted to opposite spatial frequencies respective of their tilt angles, ideally without overlap in the SFD.

### 1. Optical multiplexing

In optical multiplexing [11-30], multiple sample and reference beam pairs with different $\theta_x$ and $\theta_y$ combinations are projected onto the digital camera simultaneously, each of which creating an off-axis hologram with a different interference fringe direction that positions one wavefront in the SFD without overlapping other terms.

The simultaneous projection of all beams on the camera may create unwanted interference between nonmatching pairs. This situation can be avoided by either using a different wavelength for each sample and reference beam pair, coherence gating, or different polarization states [22]. Using these methods allows optical recording of an ideal multiplexed hologram, as described by Eq. (8).

Various basic architectures have been demonstrated for optical multiplexing. These include multiplexing two holograms by positioning two complex wavefronts in two orthogonal directions [11-21], which can be generalized to multiplexing three [22-25], four [26-28], five [29], or even six holograms [30], all without SFD overlap.

The possible applications for optical multiplexing are countless; Lohmann was the first to apply this idea [11], multiplexing the two polarization components of the electrical field in the hologram plane, thus allowing complete recording of the electrical field. This technique was later generalized by Ohtsuka et al. and Colomb et al. [12, 13]. Kühn et al. and Turko et el. multiplexed two [14,15] and three [25] holograms of the same scene with different wavelengths, for the purpose of calculating a new hologram with a much larger synthetic wavelength, preventing the phase ambiguity problem. Girshovitz at al., Frenklach et al., and Rotman-Nativ et al. multiplexed two [16, 17, 19] and three [22] different fields of view, creating a larger imaging area. Wu et al. used a complex optical system that multiplexes four fields of view using two wavelengths [26]. Chowdhury et al. [23] and Nygate et al. [18] multiplexed a digital hologram with regular fluorescence microscopy, allowing a completely registered combination of the two imaging modalities. Wang et al. multiplexed a series of three consecutive holograms of an ultrafast event of the femtosecond order from the same view angle [24]. Wolbromsky et al. multiplexed four slices within a thick sample, allowing multiple-depth imaging in a single acquisition [27]. Tian et al. multiplexed four low-resolution LED illuminations for Fourier Ptychography [28]. Finally, Dardikman et al. multiplexed two angular views of the same scene over time to enable 4-D phase unwrapping [20], and Kostencka et al. multiplexed two angular views for reducing the data acquisition time in tomographic phase microscopy [21].

### 2. Digital Multiplexing

Digital multiplexing of off-axis holograms refers to the multiplexing operation performed not by the optical setup, but rather implemented in the computer. It is used mainly for speeding up hologram reconstruction, by applying simple arithmetic operations in the hologram domain to compress multiple wavefronts into a single hologram, from which the SFD of all wavefronts can be retrieved with a single 2-D DFT [31-35].

The multiplexed holograms synthesized digitally do not consist of any unwanted interference between nonmatching pairs, as is assumed in Eq. (8), since each recording is done separately. However, the concept of sum is further generalized from the one described in Eq. (8), to a complex-weighted sum; since in digital multiplexing, the off-axis phase $\varphi_o(x, y)$ is typically identical for all holograms, the desired shift in the SFD is usually achieved by multiplying each hologram by a different complex exponent in the hologram domain prior to summation [32, 34, 35], as can be seen in Eq. (10):

$$I_{\text{multiplexed}}(x, y) = \sum_{k=1}^{N} \exp\left[ j(x \cdot u_{s,k} + y \cdot v_{s,k}) \right] \quad (10)$$
$$\cdot \left| E_{s,k}(x,y) + E_{r,k} \cdot \exp\left[ j \cdot \varphi_o(x, y) \right] \right|^2,$$

where $u_{sk}$ and $v_{sk}$ express the shift on the $u$ and $v$ axes in the SFD, respectively, for the $k$'th hologram, expressed as fractions of the respective cutoff angular frequencies $\omega_{c,u}$ and $\omega_{c,v}$.

Note that the shift achieved here is uniform for both the DC and CC terms, and is not of opposite directions, as this multiplication is equivalent to convolving the entire hologram with a shifted Dirac delta function in the SFD:

$$FT\{I_{\text{multiplexed}}(x, y)\} = \sum_{k=1}^{N} \delta[u - u_{s,k}, v - v_{s,k}] *$$
$$\left\{ \left[ FT\left\{ |E_{s,k}(x,y)|^2 \right\} + |E_{r,k}|^2 \cdot \delta(u,v) \right] \right.$$
$$+ |E_{r,k}| \cdot FT\left\{ |E_{s,k}(x,y)| \exp\left[ j \cdot \varphi_{s,k}(x,y) \right] \right\} \quad (11)$$
$$* \delta\left[ u + \frac{2\pi}{\lambda} \cdot \sin(\theta_x), v + \frac{2\pi}{\lambda} \cdot \sin(\theta_y) \right]$$
$$+ |E_{r,k}| \cdot FT\left\{ |E_{s,k}(x,y)| \exp\left[ -j \cdot \varphi_{s,k}(x,y) \right] \right\}$$
$$\left. * \delta\left[ u - \frac{2\pi}{\lambda} \cdot \sin(\theta_x), v - \frac{2\pi}{\lambda} \cdot \sin(\theta_y) \right] \right\}.$$

An alternative way of performing digital multiplexing includes summing the original hologram with a 90° rotated hologram, placing the CC terms on the orthogonal axis, thus preventing overlap in the SFD [31], yet this is limited to multiplexing two holograms. Nevertheless, this idea can be further generalized to rotating multiple holograms in small angular increments prior to summation, as has been demonstrated for up to five holograms [29]. In order for this to be relevant for speeding up reconstruction though, fast image rotation must be implemented on a dedicated hardware.

A third approach is creating a complex synthetic hologram that contains one hologram in its real part and a second hologram in its imaginary part [33], as follows:

$$I_{\text{multiplexed}}(x, y) = I_1(x, y) + j \cdot I_2(x, y). \tag{12}$$

In this manner, the CC terms of both holograms spatially overlap in the SFD, yet are completely separable. If the first overlapping CC term is $CC_1+j \cdot CC_2$ (where $CC_1$ is attributed to $I_1$ and $CC_2$ is attributed to $I_2$), then due to the linearity trait of the Fourier transform, its counterpart equals $CC_1^*+j \cdot CC_2^*$. Thus, by taking the complex conjugate of the latter, we obtain $CC_1-j \cdot CC_2$, enabling the lossless retrieval of both CC terms. Note that even though we refer to $I_1$ as the real part and to $I_2$ as the imaginary part, both holograms can actually be complex. Either way, as long as the aim of the digital multiplexing is speeding up reconstruction, the multiplexed hologram can contain complex values.

The most efficient digital multiplexing architecture, recently suggested by us, includes multiplexing sixteen wavefronts into a single complex hologram [35], but can only be applied to DC-free holograms. The most efficient architecture that does not mandate DC-free holograms was suggested by Sha et al. and allows multiplexing of eight wavefronts into a single complex hologram [34]. Both approaches use a combination of the complex exponent multiplication method [Eq. (10)] and the complex encoding method [Eq. (12)].

### C. Slightly off-axis holography

Slightly off-axis (SOA) holography is a combination of on-axis and off-axis holography [6, 36]. In this method, the off-axis concept for shifting the CC terms in the SFD by tilting one of the beams is used, but not such that the two CC terms are completely separated from the DC term, but only such that they do not overlap with one another [see Fig. 1(c)]. Thus, the spatial bandwidth requirements are lower than those of off-axis holography, and only two phase-shifted holograms (rather than three in on-axis holography) are needed for reconstruction. This can be achieved, since the sign of the CC terms in the second hologram, which is $\pi$ shifted, is opposite to their sign on the first, such that the overlapping DC term can be discarded by simple subtraction of the two holograms. Once the DC term is eliminated, one can easily isolate one of the non-overlapping CC terms from the SFD.

Similarly to off-axis holography, SOA holography leaves some free space in the SFD that can be used for either optical or digital multiplexing; Min et al., for example, optically multiplexed two slightly off-axis holograms of the same scene with different wavelengths, for the purpose of calculating a new hologram with a synthetic wavelength, preventing the phase ambiguity problem [37]; and Zhong et al. presented multiple digital multiplexing architectures for speeding up reconstruction in SOA holography [38].

## 3. Spatial bandwidth efficiency analysis

### A. Sampling theory

According to the Nyquist–Shannon sampling theorem, the cutoff frequency of the SFD is half the sampling frequency. Thus, if the detector resolution in the $x$ and $y$ directions is $\Delta x$ and $\Delta y$, respectively, the coinciding cutoff angular frequencies are given by:

$$\omega_{c,u} = \frac{2\pi}{2\Delta x} = \frac{\pi}{\Delta x}, \quad \omega_{c,v} = \frac{2\pi}{2\Delta y} = \frac{\pi}{\Delta y}. \tag{13}$$

Practically, the pixel is usually a square ($\Delta x = \Delta y$); thus the minimal cutoff frequency requirement on any axis applies to both axes.

In this analysis, we assume optical in-focus acquisition of dense samples (such as biological cells), such that the frequency content of the sample employs the entire range. Assuming that the sample maximal spatial angular frequency defined by the optical setup is $\omega_s$ on both axes, each of the CC terms occupies a bandwidth of $2\omega_s$ (as the reference beam is assumed to contain only the zero frequency), and the auto-correlation terms occupy a double spatial bandwidth of $4\omega_s$ (in coherent imaging).

The maximal spatial angular frequency of the sample that can be realized by the optical setup is defined by:

$$\omega_s = \frac{2\pi}{M \cdot d}, \tag{14}$$

where $M$ is the optical magnification and $d$ is the diffraction limited spot diameter, defined by the Abbe's criterion in coherent imaging as the ratio between the illumination wavelength and the numerical aperture [39,40]:

$$d = \frac{\lambda}{NA}. \tag{15}$$

### B. Efficient holographic imaging architectures

#### 1. On-axis holography

In on-axis holography, both the auto-correlation terms and the CC terms are centered around the zero frequency [Fig. 1(a)]. Thus, in order to capture the full frequency range of the CC terms, a cutoff frequency of only $\omega_c = \omega_s$ is needed on both axes. Since the maximal frequency of the DC term is twice that of the CC terms, the DC term does not fit in the SFD with a cutoff frequency of $\omega_s$, and is split on the two sides of the SFD according to the cyclic property of the DFT, as can be seen in Fig. 2(a). Nevertheless, since the DC term is usually not of interest, this is not significant.

#### 2. Off-axis holography

A few architectures have been previously suggested for efficiently optically multiplexing several off-axis holograms while maintaining the full frequency content of all samples without overlap.

Tahara at el. [41] proposed the space-bandwidth capacity enhanced (SPACE) architecture for the efficient recording of a single hologram, where each of the CC terms is split on the two sides of the SFD, using the cyclic property of the DFT. This method, illustrated in Fig. 2(b), uses a cutoff frequency of only $\omega_c = 2.5616\omega_s$ by choosing the CC shifts as $u_0 = 0.6096\omega_c$ and $v_0 = \omega_c$. By choosing $u_0 = v_0 = 0.6796\omega_c$, the CC terms can be placed on the diagonal, enabling the compression of up to two holograms with the use of a slightly higher cutoff frequency of $\omega_c = 3.1213\omega_s$ [41]. This architecture is illustrated in Fig. 2(c).

Recently, we have suggested six-pack holography (6PH), in which six off-axis holograms with six different fringe orientations are multiplexed on the same camera plane. This architecture uses a cutoff frequency of $\omega_c = 4\omega_s$ [30], as illustrated in Fig. 2(d). 6PH represents the optimal bandwidth consumption when the DC terms are present.

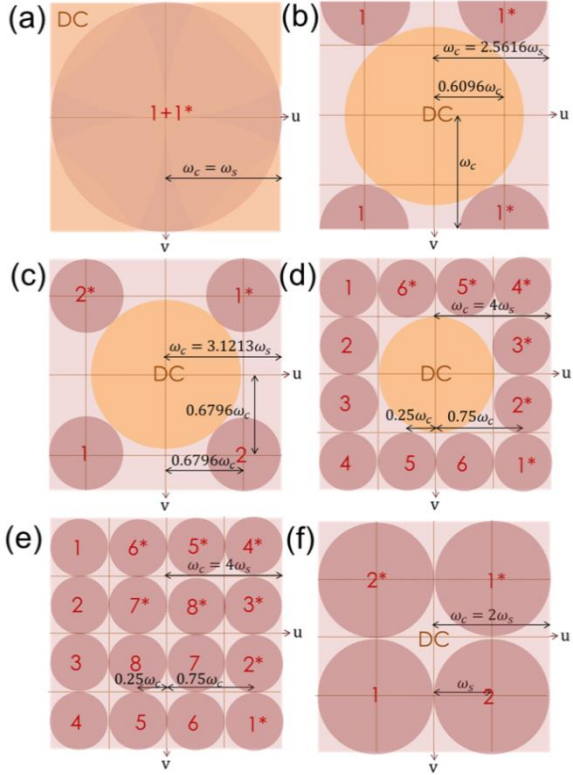

Fig. 2. Schematic illustrations of the SFD power spectra for various spatial bandwidth-efficient holographic imaging architectures, including bandwidth calculations, assuming the same number of camera pixels. (a) Optimal on-axis holography. (b) SPACE. (c) Diagonal off-axis multiplexing. (d) 6PH. (e) 8PH with the DC terms removed. (f) Diagonal slightly off-axis multiplexing with the DC terms removed. DC denotes the auto-correlation terms and the numbered circles around it denote the CC terms, where coinciding complex conjugate CC terms are denoted by the same number with and without an asterisk.

We have later also suggested multiplexing eight holograms (8PH) [35], by utilizing the space occupied by the DC terms. This architecture is illustrated in Fig. 2(e). The 8PH architecture can be applied optically in the general case by removing the DC terms, as can be done by acquiring two phase-shifted holograms, similarly to slightly off-axis holography [6, 35].

*3. Slightly off-axis holography*

In slightly off-axis holography, the CC terms are slightly shifted such that they do not overlap with each other, but each of them overlaps with the DC terms, as demonstrated in Fig. 1(c). Thus, in order to capture the full frequency range of the CC terms, a cutoff frequency of $\omega_c = 2\omega_s$ is needed on the $u$ axis, and a cutoff frequency of $\omega_c = \omega_s$ is needed on the $v$ axis. However, assuming that the detector pixel size is identical on both axes, as is usually the practical case, the effective minimal cutoff frequency is $\omega_c = 2\omega_s$ for both axes. For that case, by placing the CC terms on the diagonal, two holograms can be efficiently multiplexed, as can be seen in Fig. 2(f).

### C. Quantitative spatial-bandwidth efficiency comparison

We now compare the efficiency of the spatial-bandwidth consumption for the different architectures discussed in Section 3B and illustrated in Fig. 2. We define the efficiency score as follows:

$$E_f = \frac{\omega_s}{\omega_c} \cdot \frac{N_w}{N_a}, \quad (16)$$

where $\omega_c$ is the cutoff angular frequency required for the method, $\omega_s$ is the maximal angular frequency of the sample, $N_w$ is the number of wavefronts encoded in the method, and $N_a$ is the number of acquisitions needed for reconstruction. This score expresses the ratio between the effective number of wavefronts per acquisition ($N_w / N_a$) and the relative cutoff angular frequency required for capturing the full frequency range of all samples without overlap ($\omega_c / \omega_s$). The calculation of the efficiency score for each of the architectures is given in Table 1.

As can be seen from Table 1, the 6PH architecture is the most efficient holographic method in terms of number of wavefronts per bandwidth, whereas the optimal on-axis holography is the least efficient. Nevertheless, other considerations should be noted.

In order to compare the different methods, we assumed that two acquisitions are needed for reconstruction both in 8PH and in diagonal SOA multiplexing. Practically, though, acquiring these two phase-shifted frames can be achieved simultaneously by using both exits of a Mach-Zehnder interferometer [6, 35], rather than by placing wave plates or a piezo mirror and sequentially acquiring additional images, as is usually the case with the three frames needed for on-axis holography. This is an important consideration when imaging dynamic samples, since in this case the cost of multiple acquisitions may be an unwanted change in the sample during acquisitions, thus one should prefer to use methods that require only a single acquisition, such as SPACE, diagonal off-axis or 6PH; however, since we can simultaneously acquire two phase shifted holograms using two synchronous cameras, effectively acting as a single acquisition (if we ignore camera registration problems), 8PH could be considered to be the most efficient method for dynamic samples, with an effective efficiency score of 2. Note that even if we consider a simultaneous acquisition implementation for on-axis holography, such as the one suggested by Awatsuji et al. [7], effectively acting as a single acquisition, we would still get an effective efficiency score of only 1, still placing on-axis holography as considerably less bandwidth-efficient in comparison to 6PH and 8PH.

Another consideration that has to be taken into account is reduction in SNR when spatially multiplexing several holograms, due to sharing the dynamic range of the camera gray scale levels; this is discussed in detail in the following section.

**Table 1. Comparison of various digital holography architectures. $\omega_c$ is the cutoff angular frequency required, $\omega_s$ is the maximal angular frequency of the sample, $N_w$ is the number of wavefronts encoded, $N_a$ is the number of acquisitions required for reconstruction, and $E_f$ is the efficiency score defined in Eq. (16).**

|  | $\omega_c/\omega_s$ | $N_w$ | $N_a$ | $E_f$ |
|---|---|---|---|---|
| Optimal on-axis | 1 | 1 | 3 | 0.33 |
| SPACE | 2.56 | 1 | 1 | 0.39 |
| Diagonal off-axis | 3.12 | 2 | 1 | 0.64 |
| 6PH | 4 | 6 | 1 | 1.5 |
| 8PH | 4 | 8 | 2 | 1 |
| Diagonal SOA | 2 | 2 | 2 | 0.5 |

### 4. Reconstruction quality analysis

To quantify the quality of the reconstructed image in the presence of shot noise for the various holographic architectures discussed, we conducted numerical simulations. We assumed the wavelength to be $\lambda = 633$ nm, the camera pixel size to be $\Delta x = \Delta y = 5.2$ μm and the objective numerical aperture to be NA=1.4.

First, we defined a 3-D refractive index (RI) test target, as can be seen in Fig. 3(a). This test target, meant to imitate a biological cell suspended in water, has a major axis radius of 6 μm and a minor axis radius of 3 μm.

For each of the architectures, the proper magnification was calculated from the defined optical parameters using Eqs. (13)-(15) to achieve the precise ratio between the sample bandwidth and the cutoff frequency shown in Fig. 2. Then, the respective phase and amplitude profiles were calculated; the sample phase profile [Fig. 3(b)] was calculated from the 3-D RI distribution as:

$$\varphi(i,j) = \frac{2\pi}{\lambda} \Delta k \sum_{k=1}^{N} n(i,j,k), \tag{17}$$

where $\Delta k$ is the grid element length in the $k$ dimension, $N$ is the grid size in the $k$ dimension ($N = 256$ here), and $n(i,j,k)$ is the 3-D RI distribution; the sample amplitude profile [Fig. 3(c)] was estimated from the phase profile by normalizing the phase values and subtracting them from a constant matrix, followed by replacing the object outer edge value with a lower constant, to imitate diffraction effects. Next, the sample wavefront was constructed using the above amplitude and phase profiles, and filtered in the SFD using a low pass filter with a suitable cutoff frequency $\omega_s$ [Eq. (14)], to account for the diffraction limit present in an actual optical recording due to the optical setup. The final phase and amplitude profiles used as the ground truth for comparison were then extracted from this filtered wavefront, as can be seen in Figs. 3(d) and 3(e), respectively.

The reference wavefront, which is assumed to have constant phase and amplitude, was calculated by inputting a matrix of ones as $n(i,j,k)$ to Eq. (17). The holograms for all architectures discussed in Section 3B were then generated using Eqs. (2), (5), (7) and (8), with their simulated SFDs shown in Fig. 4. In architectures that include multiplexing several wavefronts, the same wavefront was used numerous times.

After the holograms were generated, they were all stored in an 8-bit image format, assuming an ideal detector which can set the lowest intensity value to 0 and the maximum value to 255. To simulate a realistic recorded intensity, shot noise (Poisson type) was then introduced on each image. The final holograms are presented in Fig. 5.

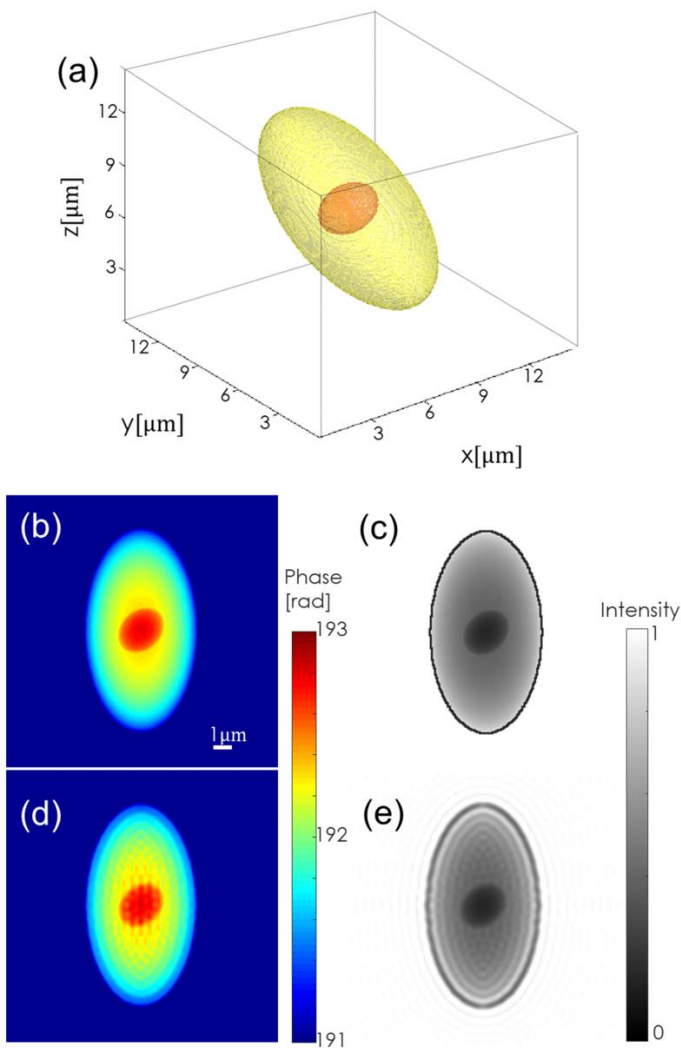

Fig. 3. Numerical simulation inputs. (a) Test target 3-D RI distribution. Yellow indicates an RI value of 1.35, and red indicates an RI value of 1.37. (b) Original phase image. (c) Original amplitude image. (d) Filtered phase image. (e) Filtered amplitude image.

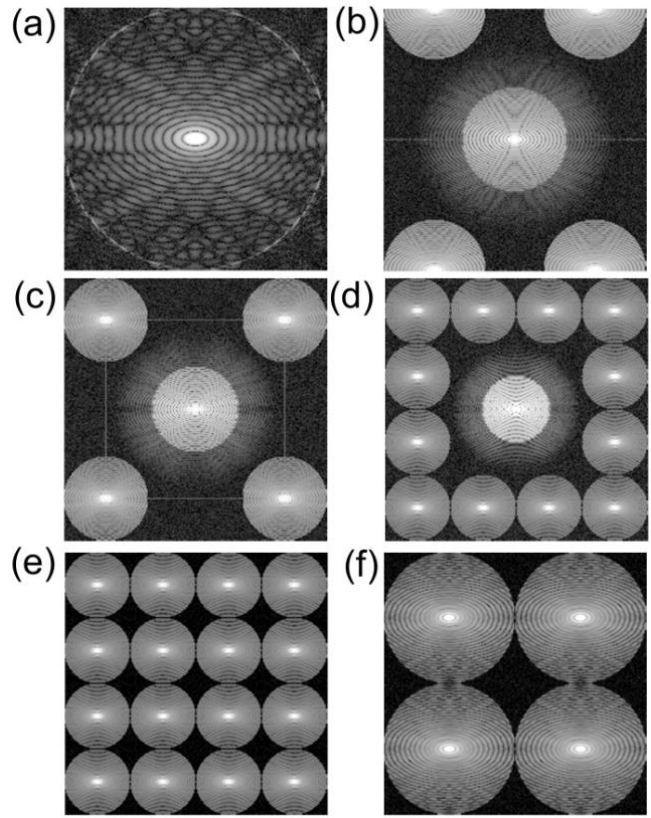

Fig. 4. Numerical simulation of the SFD power spectra for various spatial bandwidth-efficient holographic imaging architectures. (a) Optimal on-axis holography. (b) SPACE. (c) Diagonal off-axis multiplexing. (d) 6PH. (e) 8PH. (f) Diagonal slightly off-axis multiplexing. For (e) and (f), the final SFD, obtained after subtracting the two phase-shifted holograms, is presented.

In order to understand the characteristics and pitfalls of the 6PH and 8PH architectures, we also implemented a simulation imitating standard off-axis holography, without multiplexing, as shown in Fig. 1(b), with a layout of $\omega_c = 4\omega_s$.

The phase and amplitude profiles were then reconstructed from the off-axis and slightly off-axis holograms using the Fourier filtering method [31], and from the on-axis holograms using simple arithmetic operations applied on the three phase-shifted holograms in the image domain [Eq. (3)]. In the cases of diagonal SOA holography and of 8PH, the two phase-shifted holograms were subtracted prior to applying the reconstruction algorithm, to eliminate the overlapping DC components.

In the Fourier filtering method, we first apply 2-D DFT on the hologram; then we locate and crop the relevant (non-conjugate) CC terms in the SFD, each with a window size of $2\omega_s \times 2\omega_s$, and apply an inverse 2-D DFT to each of them; finally, we decompose each extracted wavefront to amplitude and phase, where the phase undergoes 2-D phase unwrapping for solving $2\pi$ ambiguities using an algorithm based on sorting by reliability following a noncontinuous path [42]. Once this process is complete, all reconstructed profiles are resized back to the original image dimensions of 256×256 pixels by interpolation. In cases where several wavefronts were multiplexed, all were reconstructed.

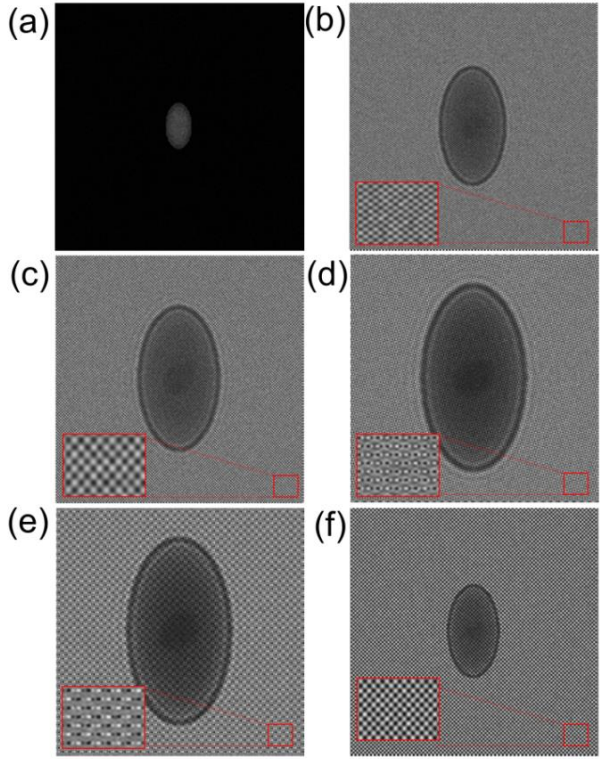

Fig. 5. Simulated holograms for the various architectures. (a) Optimal on-axis holography. (b) SPACE. (c) Diagonal off-axis holographic multiplexing. (d) 6PH. (e) 8PH. (f) Diagonal slightly off-axis holographic multiplexing. For (a), (e) and (f), several phase-shifted holograms were generated as needed for reconstruction, but only one is shown. Note that different magnifications were required for the different spatial bandwidth-efficient architectures for optimal usage of the spatial bandwidth. The red rectangle shows a close up image of the interference fringes.

Due to the scalable nature of amplitude, the quality of the reconstructed amplitude images could not be properly evaluated quantitatively. Instead, the reconstructed images are presented in Fig. 6, relative to the corresponding ground truth amplitude images.

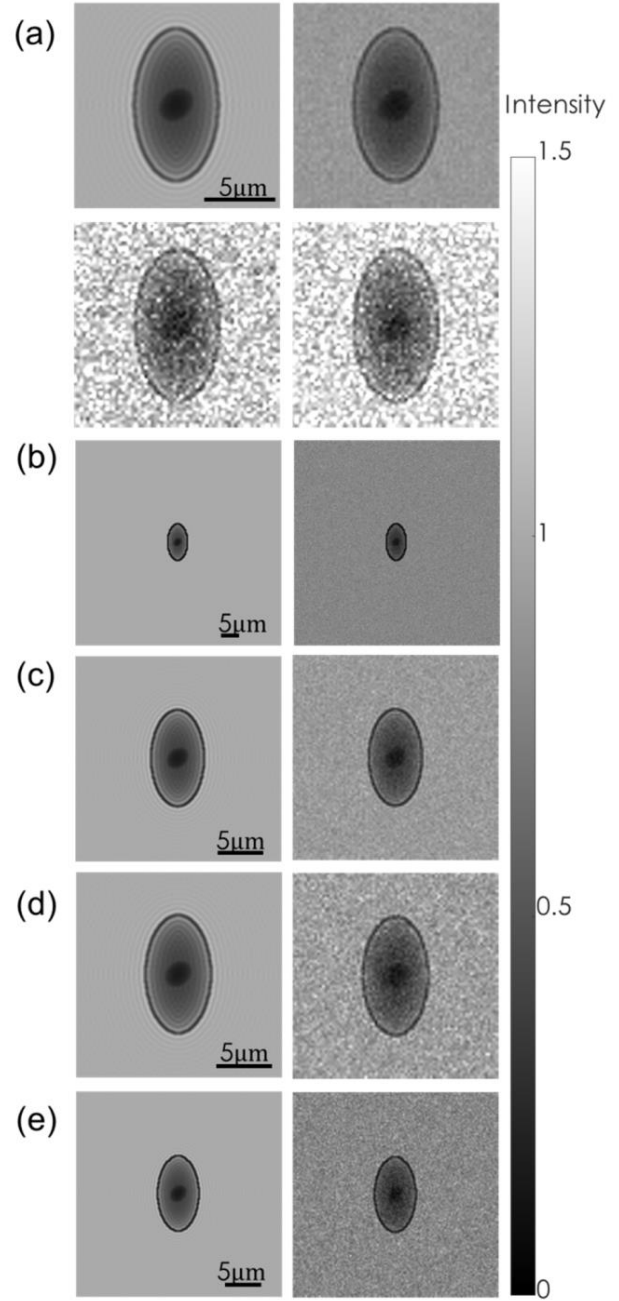

Fig. 6. Amplitude reconstruction results for an 8-bit ideal detector. (a) Various off-axis holographic architectures using $\omega_c = 4\omega_s$; top left: ground truth; top right: non-multiplexed off-axis holography; bottom left: 6PH; bottom right: 8PH. (b) Optimal on-axis holography; left: ground truth; right: reconstruction (c) SPACE; left: ground truth; right: reconstruction. (d) Diagonal off-axis holographic multiplexing; left: ground truth; right: reconstruction. (e) Diagonal slightly off-axis holographic multiplexing; left: ground truth; right: reconstruction. For architectures where several wavefronts are multiplexed, only the first reconstruction is displayed.

As can be seen from Fig. 6(a), the quality of the reconstruction clearly degrades due to multiplexing numerous wavefronts, attributed to the limited dynamic range of the camera. Generally, the reconstruction of amplitude obtained from holograms that contain only one wavefront, including the on-axis holography [Fig. 6(b)] and SPACE [Fig. 6(c)] seems superior to the one obtained from holograms that contain two wavefronts, including the diagonal off-axis [Fig. 6(d)] and slightly off-axis [Fig. 6(e)] holographic architectures.

In order to quantify the quality of the reconstructed phase images relative to the ground truth phase images, we used two types of metrics; the first and most naive one is the mean squared error (MSE); whereas the second one is the signal to noise ratio (SNR), based on computing the ratio between the summed squared magnitude of the ground truth phase image to that of the noise, given by the absolute difference between the ground truth phase image and the reconstructed phase image.

Both metrics were used to estimate the phase reconstruction results obtained using all discussed holographic architectures. In order to account for samples with low amplitude modulation, such as most biological cells *in vitro*, a scenario of uniform amplitude modulation was also considered. The results of both scenarios are given in Table 2 and in the second and fifth columns of Fig. 7, with the corresponding ground truths in the first and fourth columns. The difference in the ground truth phase profile with and without amplitude modulation is due to the low pass filter applied on the sample wavefront to simulate the resolution limit of the optical system. Note that in order to make the comparison between the different architectures as fair as possible, each image was cropped to contain only the minimal rectangular bounding box surrounding the object prior to the calculations, such that architectures with low magnifications (and thus many background pixels) will not get a disproportionate advantage over those with large magnifications. All results were also averaged over 150 simulations each, with the noise profile randomly generated for each simulation. For architectures where multiple wavefronts are multiplexed, the final values per simulation were calculated as the average.

As can be seen from Table 2, the standard off-axis architecture takes the lead when an 8-bit detector is used, both in terms of SNR and MSE, either with or without amplitude modulation; for both cases the 6PH and 8PH methods present the worst results, presumably as a result of dynamic range sharing by the multiple wavefronts.

Comparing the two simulated scenarios, it is visible that without amplitude modulation the reconstruction quality improves for all architectures, in agreement with recent research [43], stating that for multiplexed off-axis holograms the phase error is more dominant in areas where the amplitude value is low (i.e. areas with higher amplitude modulation).

**Table 2. Phase reconstruction quality estimation for an 8-bit ideal detector. The green marker highlights the best method in each column, and the red highlights the worst.**

|  | With amplitude modulation | | Without amplitude modulation | |
| --- | --- | --- | --- | --- |
|  | SNR [dB] | MSE [rad²] | SNR [dB] | MSE [rad²] |
| Optimal on-axis | 12.55 | 0.0429 | 19.88 | 0.0077 |
| SPACE | 18.00 | 0.0120 | 21.34 | 0.0053 |
| Diagonal off-axis | 15.13 | 0.0228 | 17.96 | 0.0117 |
| Standard off-axis | 19.93 | 0.0074 | 22.00 | 0.0046 |
| 6PH | 9.98 | 0.0737 | 11.28 | 0.0543 |
| 8PH | 10.77 | 0.0613 | 11.72 | 0.0491 |
| Diagonal SOA | 13.66 | 0.0338 | 18.51 | 0.0103 |

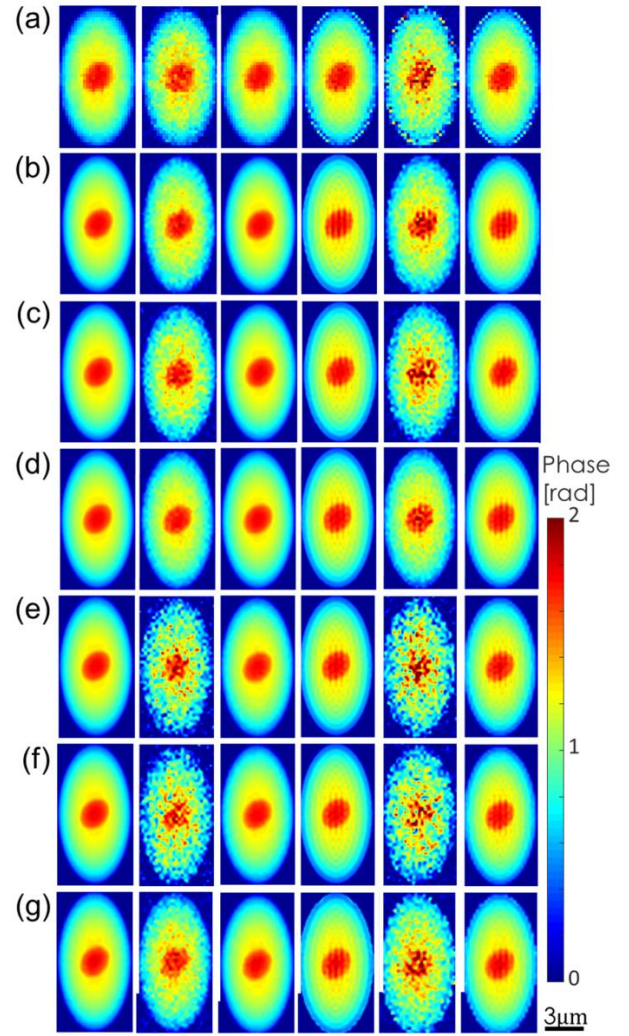

Fig. 7. Phase reconstruction results. (a) Optimal on-axis holography. (b) SPACE. (c) Diagonal off-axis holographic multiplexing. (d) Non-multiplexed off-axis holography. (e) 6PH. (f) 8PH. (g) Diagonal slightly off-axis holographic multiplexing. First column to the left: ground truth without amplitude modulation. Second column: reconstruction without amplitude modulation for an 8-bit ideal detector. Third column: reconstruction without amplitude modulation for a 16-bit ideal detector. Fourth column: ground truth with amplitude modulation. Fifth column: reconstruction with amplitude modulation for an 8-bit ideal detector. Sixth column: reconstruction with amplitude modulation for a 16-bit ideal detector. For architectures where several wavefronts are multiplexed, only the first reconstruction is displayed. The phase maps we chose to present here have MSE values close to the average, as seen in Tables 2 and 3.

It is interesting to note that 8PH presents slightly better results than 6PH both in terms of SNR and MSE, where 8PH is actually expected to be worse due to multiplexing more wavefronts. This may be attributed to the subtraction of phase-shifted holograms needed for reconstruction, possibly causing some of the error factors to be subtracted, thus improving reconstruction. This innate advantage, also relevant to diagonal SOA holography and on-axis holography, can also be noticed visually in Fig. 4, presenting the SFD for all architectures prior to introducing shot noise. Figures 4(e) and 4(f), presenting the SFDs of subtracted holograms, clearly show a cleaner signal than all others.

Note that the different magnifications used for the different architectures may also affect the results due to numerical errors, possibly degrading reconstruction quality in methods using lower magnifications, such as on-axis holography.

To verify that the low SNR and high MSE obtained using the 6PH and 8PH architectures is indeed a result of dynamic range limitations, we also ran this simulation for an ideal 16 bit detector, with the results shown in Table 3 and in the third and sixth columns of Fig. 7.

**Table 3. Phase reconstruction quality estimation for a 16-bit ideal detector, averaged over 150 simulations each. The green marker highlights the best method in each column, and the red highlights the worst.**

|  | With amplitude modulation | | Without amplitude modulation | |
| --- | --- | --- | --- | --- |
|  | SNR [dB] | MSE [rad$^2$] | SNR [dB] | MSE [rad$^2$] |
| Optimal on-axis | 37.57 | 1.32×10$^{-4}$ | 43.43 | 3.41×10$^{-5}$ |
| SPACE | 21.88 | 0.0049 | 25.51 | 0.0020 |
| Diagonal off-axis | 22.04 | 0.0046 | 24.61 | 0.0025 |
| Standard off-axis | 21.97 | 0.0046 | 23.80 | 0.0030 |
| 6PH | 21.78 | 0.0048 | 23.55 | 0.0032 |
| 8PH | 21.82 | 0.0048 | 23.58 | 0.0032 |
| Diagonal SOA | 22.27 | 0.0046 | 27.21 | 0.0014 |
| Optimal on-axis with 10% error in phase shift | 21.25 | 0.0057 | 21.30 | 0.0056 |
| 8PH with 10% error in phase shift | 21.76 | 0.0049 | 23.57 | 0.0032 |
| Diagonal SOA with 10% error in phase shift | 22.26 | 0.0046 | 27.19 | 0.0014 |

As can be seen in Table 3, when using a camera with a larger bit depth, the image corruption caused by multiplexing several wavefronts becomes negligible. This is demonstrated by the fact that the result obtained by using standard off-axis holography is nearly identical to the ones obtained using either 6PH or 8PH. Interestingly, in this case, the reconstruction quality using most architectures is similar, excluding on-axis holography that improves dramatically, taking the lead above all others.

Although on-axis holography defeated all other holographic methods in reconstruction quality for the 16-bit detector, this simulative comparison did not consider other possible error factors in on-axis holography, such as sample change between acquisitions (which is relevant for sequential acquisition) and errors caused by inaccurate phase shifts between the holograms. These additional error factors may also exist in 8PH and in diagonal SOA. To assess the influence of such error factors on the final reconstruction, we introduced a 10% error in phase-shift for on-axis holography, 8PH and diagonal SOA, with the results presented in the bottom three lines of Table 3. As can be seen from these results, the apparent relative advantage of on-axis holography over the other methods is quickly lost by a 10% deviation in the experimental phase shift, while 8PH and diagonal SOA are barely affected.

Finally, note that while this analysis shows the general trend in SNR, it does not consider many real-life phenomena affecting hologram optical acquisition, such as speckle noise caused by the high coherence length of the laser, readout noise, or some practical problems of a non-ideal detector.

## 5. Conclusions

In this paper, we reviewed the most efficient possible architectures for digital holographic imaging considering SFD multiplexing, and analyzed them both in terms of spatial bandwidth efficiency and of reconstruction quality. All architecture analyzed here assume optical in-focus acquisition of dense samples, such that the frequency content of the sample employs the entire range, thus not using any sparse representations that could possibly improve compression ratio [44].

For the spatial bandwidth consumption analysis, we defined an efficiency score, based on the ratio between the effective number of wavefronts per acquisition and the relative cutoff frequency needed for capturing the full frequency range of all wavefronts without overlap, and used it to analyze the various possible holographic architectures and compare their efficiencies. Based on this score, we found that the off-axis 6PH architecture is the most efficient method, and the optimal on-axis holography is the least efficient. We noted that this score is imperfect, as it is unable to consider the fact that when only two phase-shifted holograms are needed, they can be acquired simultaneously by using both exits of a Mach-Zehnder interferometer, and thus may be effectively considered as a single acquisition, placing 8PH as the most efficient method. If two simultaneous acquisitions are allowed, 8PH would obviously also be favorable in a scenario where multiple dynamic wavefronts need to be captured simultaneously.

For the reconstruction quality analysis, we performed numerical simulations imitating all discussed architectures in the presence of shot noise, and compared the quality of the results both in terms of MSE and SNR. In this analysis, we found that for an 8-bit detector the most bandwidth-efficient methods – 6PH and 8PH – supply the lowest quality reconstructions, presenting a clear tradeoff between efficiency and quality. Nevertheless, when using a 16-bit detector and considering additional error factors, this issue becomes negligible, and all architectures present similar reconstruction quality. Thus, we conclude that in cases where bandwidth efficiency is of the highest priority, 6PH and 8PH are good choices, as long as a detector with a large bit depth is used.

**Funding Information.** Horizon 2020 European Research Council (ERC) 678316; Tel Aviv University Center for Light-Matter Interaction.